\documentclass[proof]{WileyASNA-v1}

\articletype{Article Type}%

\received{24 December 2020}
\revised{31 December 2020}
\accepted{1 January 2021}

\begin{document}

\title{Studying the onset of deconfinement with multi-messenger astronomy of neutron stars}

\author[1,2,3]{David Blaschke*}
\author[1]{Mateusz Cierniak**}

\authormark{David Blaschke \& Mateusz Cierniak}

\address[1]{\orgdiv{Institute of Theoretical Physics}, \orgname{University of Wroclaw}, \orgaddress{50-204 Wroclaw, \country{Poland}}}

\address[2]{\orgdiv{Bogoliubov Laboratory of Theoretical Physics, JINR Dubna, 141980 Dubna, Russia}}  

\address[3]{\orgdiv{National Research Nuclear University (MEPhI), 115409 Moscow, Russia}}  

\corres{* \email{david.blaschke@uwr.edu.pl}\\ ** \email{mateusz.cierniak@uwr.edu.pl} }

\abstract{
With the first multi-messenger observation of a binary neutron star merger (GW170817) new constraints became available for masses and radii of neutron stars. 
We introduce a class of hybrid EoS that fulfils all these constraints and predicts a region in the mass-radius diagram that could be populated only by hybrid neutron stars with quark matter cores. 
A confirmation of this conjecture would be provided when the NICER radius measurement for the high-mass pulsar PSR J0740+6620 yields a radius significantly less than 11 km.
Would this radius measurement yield a result in excess of 12 km, this would allow for both, a purely hadronic and a hybrid 
nature of this star.
In the latter case the maximum mass could reach $2.6~M_\odot$ so that  the lighter object in the asymmetric binary merger GW190814 %with 2.6 $M_\odot$ 
could have been a hybrid star. 
We demonstrate that this high mass can be compatible with an early onset of deconfinement for stars with masses below $1~M_\odot$ and the occurrence of low-mass twin stars.
In such a case the remnant of GW170817 could be a long-lived hypermassive pulsar.
}


\maketitle

\section{Introduction}

One of the most intriguing questions in the astrophysics of neutron stars (NS) is whether their interiors harbour deconfined quark matter and how to prove or disprove such claims.  
We want to revisit this question in view of constraints on masses and radii of pulsars in the era of multi-messenger Astronomy when in the course of a binary neutron star merger event the limits for densities, temperatures and rotation frequencies of compact stars are reached.

As a heuristic tool for such an investigation, we will use in this work the fact that for hybrid star equations of state (EoS) there exists a special point in the mass-radius ($M-R$) diagram which is largely independent of the hadronic EoS~\citep{Yudin:2014mla}.
Varying the parameters of the quark matter EoS, a region of its possible location in the M-R diagram has been mapped out~\citep{Cierniak:2020eyh} which 
indicates the region where hybrid NS can be found.
We discuss that the softest purely hadronic EoS that can be considered realistic in view of existing NS constraints and that could be used to estimate the lower limit for the radius of purely hadronic NS is that of~\citep{Yamamoto:2017wre} and not that of~\citep{Akmal:1998cf}, hereafter abbreviated as APR, which has a severe hyperon puzzle problem. 
%%%%%%%%%%%%%%%%%%%%%%%%%%%%%%%%%%%%%%%%%%%

On this basis we draw these conclusions:
\begin{enumerate}
\item
We identify a first region in the $M-R$ plane which can only be accessed by hybrid NS. 
If the radius of a $2~M_\odot$ NS turns out to be less than $\sim 11.5$ km (e.g., as a result of the NICER radius measurement on the millisecond pulsar 
PSR J0740+6620~\citep{Cromartie:2019kug}, one can conclude that this NS harbors an extended quark matter core\footnote{By a different method, the authors of ~\citep{Annala:2019puf} came to the conclusion that all NS with a mass of $2~M_\odot$ should have an extended quark matter core. 
But only if the conformal bound for the squared speed of sound ($c_s^2\le 1/3$) would be violated the radius at $2~M_\odot$ could be less than 12 km.}. 
Such a measurement would exclude the "two-families" scenario~\citep{Drago:2020gqn} according to which stars with large masses $M\ge 2~M_\odot$ should be strange stars with radii $\ge 14$ km while the maximum mass of the smaller hadronic stars stays well below the $2~M_\odot$ limit.  
\item
Very massive objects such as the $2.6~M_\odot$ companion of the $23~M_\odot$ black hole in recently discovered merger GW190814~\citep{Abbott:2020khf} could be explained has hybrid neutron stars, even on a third family branch. 
The existence of such massive NS would then not exclude the existence of mass twin stars, opposite to the recent claim by~\citep{Christian:2020xwz}.
\item 
It is unlikely that these two statements are simultaneously true.
If NICER measures a NS radius at $2~M_\odot$ smaller than $\sim 10.5$ km with sufficient accuracy, the upper limit of the NS maximum mass is 
$\lesssim 2.5~M_\odot$.
This would entail that GW190814 was a binary black hole merger.
\item
We identify a second region in the $M-R$ plane which can be accessed only by hybrid NS sequences. 
It is the mass range below $\sim 1.5~M_\odot$, where radii below $\sim12.5$ km would indicate a softening of the EoS due to a phase transition.
The recent analysis by~\citep{Capano:2019eae} of the combined multi-messenger observations of the binary NS merger GW170817 \citep{TheLIGOScientific:2017qsa} suggests that $R_{1.4M_\odot}=11.0^{+0.9}_{-0.6}$ km and may thus be considered as indication for an early onset of deconfinement in neutron stars below the typical binary radio pulsar mass.
\end{enumerate}
%%%%%%%%%%%%%%%%%%%%%%%%%%%%%%%%%%%%%%%%%%%%%%%%%%%%%%%%% 

\section{A special point for hybrid stars}

\subsection{Hybrid star EoS and $M-R$ diagram}

In the absence of a unified description of cold hadronic and quark matter with a phase transition at high densities, the hybrid star matter equations of state (EoS) must be constructed by matching separate EoS for the pressure that use hadrons ($p_h(\mu)$) or deconfined quarks ($p_q(\mu)$) as degrees of freedom.  The EoS with the higher pressure is the one that is preferred by Nature and describes the bulk thermodynamics. A crossing of both alternative EoS at the critical chemical potential $\mu_c$ marks a first-order phase transition fulfilling the condition 
$p_h(\mu_c)=p_q(\mu_c)$ for the Maxwell construction.

The first of two hadronic EoS dscussed in this study is chosen from the class of relativistic desity-functional (RDF) models based 
on the ``DD2'' parametrization~\citep{Typel:2009sy} with excluded volume effects modelled according to~\citep{Typel:2016srf}. 
It describes the properties of nuclear matter at low densities up to and slightly above nuclear saturation. 
The class of DD2 models with excluded nucleon volume will mostly be used in the context of an early onset deconfinement with large latent heat and the possiblity of observing the so-called "mass twin" phenomenon.

For the deconfined quark phase, a class of constant speed of sound (CSS) quark matter models will be used. The equation of state of this model is built starting from an assumption on the relation between thermodynamic pressure and energy density (see also the Appendix of Ref.~\citep{Alford:2013aca}), 
\begin{equation} \label{css1}
    \varepsilon = \varepsilon_0+c_s^{-2} p,
\end{equation}
where $\varepsilon_0$ is a constant energy density shift and $c^2_s$ is the square of the speed of sound (also assumed to be constant). Furthermore, we define the baryonic chemical potential as
\begin{equation}
    \mu(p)=\mu_0 \left(\frac{p+B}{A}\right)^{c_s^2/(1+c_s^2)},
\end{equation}
or alternatively we express the pressure as a function of baryochemical potential
\begin{equation} 
\label{css3}
    p(\mu)=A(\mu/\mu_0)^{1+c_s^{-2}}-B,
\end{equation}
where the model parameters $A$, $B$ and $c_s^2$ are constant and $\mu_0=1$ GeV. 
The baryon density $n$ follows from the canonical relation
\begin{equation}
    \frac{\partial p(\mu)}{\partial\mu}=n(\mu)=(1+c_s^{-2})A(\mu/\mu_0)^{c_s^{-2}}.
\end{equation}
Using the above, we arrive at the energy density
\begin{equation}
    \varepsilon=\mu n-p=B + c_s^{-2} A(\mu/\mu_0)^{1+c_s^{-2}}.
\end{equation}
The relation between pressure and energy density takes the form
\begin{equation} \label{css6}
   p = c_s^2  \varepsilon - (1+c_s^2)B.
\end{equation}
From Eq.~\ref{css6} we see, that $\varepsilon_0=(1+c_s^{-2})B$. 
The pressure slope parameter $A$ does not affect the relation between pressure and energy density, but its values should be limited to a range that produces a non-negative density jump at the phase transition.
It has been shown by~\citep{Zdunik:2012dj} that the above CSS model fits well the EoS of color superconducting quark matter in the CFL phase that was obtained from a self-consistent solution of the three-flavor NJL model in the mean field approximation~\citep{Blaschke:2005uj,Klahn:2013kga}. 
%%%%%%%%%%%%%%%%%%%%%%%%%%%%%%%%%%%%%%%%%%%%%%%

\begin{figure}[htb]
\centering
\includegraphics[width=0.45\textwidth]{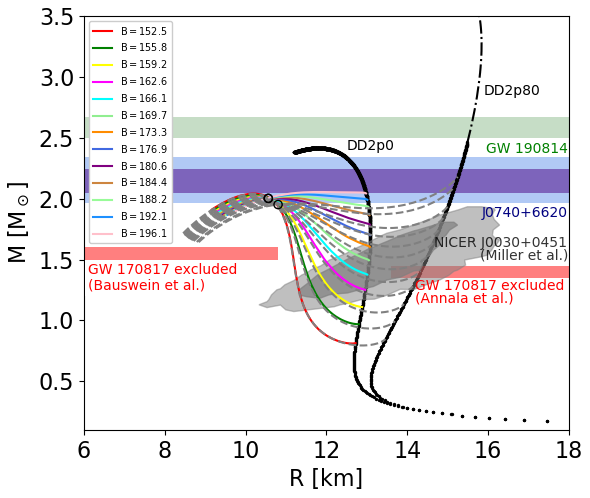} 
\caption{Mass--radius relations of neutron stars for two sets of hybrid EoS, where  the quark matter phase is described by the CSS model and the hadronic one by the EoS DD2p0 or DD2p80 (nomenclature according to~\citep{Typel:2016srf}). The values of the bag pressure $B$ in the CSS model are given in MeV/fm$^3$.
The positions of the special points marked by black circles are only marginally dependent on the hadronic phase EoS. 
The blue and grey bands correspond to $68.3\%$ and $95.4\%$ credibility intervals of the Shapiro--delay measurement of pulsar's PSRJ0740+6620 mass~\citep{Cromartie:2019kug} and the simultaneous mass-radius measurement on PSR J0030+0451 from one of the analysis teams in the NICER Collaboration~\citep{Miller:2019cac}, respectively. Red bands are regions excluded by the analysis of the signal from the neutron star merger event GW170817 according to~\citep{Bauswein:2017vtn} and~\citep{Annala:2017llu}, see also~\citep{Cierniak:2020eyh}.} 
\label{fig1}
\end{figure}

With this setting for the classes of hadronic and quark matter EoS, the hybrid star EoS are obtained by the Maxwell construction as described above.
Varying the bag pressure parameter $B$ one obtains a family of hybrid EoS that after solving the Tolman-Oppenheimer-Volkoff equations results in the 
family of sequences shown in Fig. \ref{fig1} which exhibit a special point (SP) through which they all pass.
This phenomenon has been investigated more in detail by~\citep{Yudin:2014mla} and recently also by~\citep{Cierniak:2020eyh} for different classes of quark matter 
EoS. Note that the position of the SP is only marginally dependent on the choice of the hadronic EoS, which is also demonstrated in Fig. \ref{fig1}.
For the stiffer hadronic EoS DD2p80 with the larger excluded volume parameter of the nucleon, 
a branch of unstable star configurations occurs immediately after the deconfinement transition. 
This unstable branch disconnects the purely nucleonic NS sequence from the one of stable hybrid stars. 
Such a situation is characteristic for the occurrence of a third family of compact stars~\citep{Gerlach:1968zz} which entails the existence of mass twins
\citep{Glendenning:1998ag}.
Note that all hybrid star sequences based on the DD2 EoS fulfil the observational constraints while for the stiffer DD2p80 EoS the mass for the onset of 
deconfinement should be lower than the one of the lighter neutron star in GW170817, i.e. $M_{\rm onset}<1.35~M_\odot$.
%%%%%%%%%%%%%%%%%%%%%%%%%%%%%%%%%%%%%%%%%%%%%%%%%%%%%

\subsection{Region of hybrid stars in the $M-R$ plane}

\begin{figure}[htb]
 \centering
\includegraphics[width=0.45\textwidth]{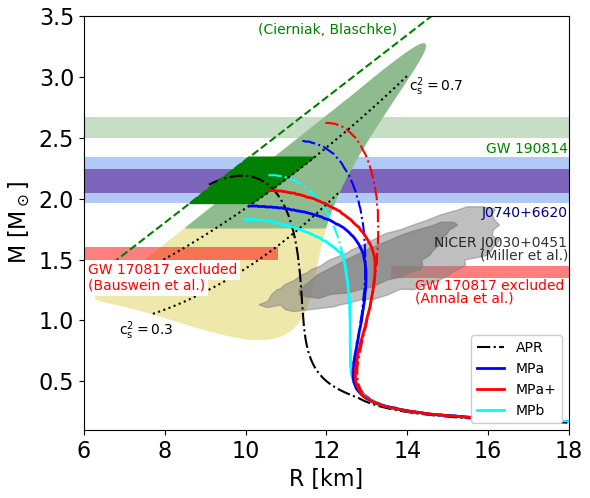} 
\caption{The SP range allowed (light green), and excluded (yellow) by the 2 $M_\odot$ constraint. 
The green dashed line shows the upper limit of masses that can be reached for a given radius of the hybrid star configurations.
The {red}, blue and cyan lines show the M-R relation for the hadronic EoS that are obtained with realistic baryon-baryon two-body and three-body interaction augmented by a repulsive short-range multi-pomeron (MP) exchange potential according to~\citep{Yamamoto:2017wre,Yamamoto:2015lwa}. 
The solid lines represent EoS parametrizations with hyperons, while dash-dotted lines represent EoS with only non-strange baryons.
The black dash-dotted line is obtained for the APR EoS~\citep{Akmal:1998cf}, also without hyperons. 
The dark green rhombic region can be assessed only by hybrid stars, see also~\citep{Cierniak:2020eyh}.} 
\label{fig2}
\end{figure}

We have explored in~\citep{Cierniak:2020eyh} the region in the $M-R$ diagram that can be accessed by hybrid neutron star configurations with quark matter cores.
As a proxy for that region, one may consider instead the area where the SP can be located. It is shown as the yellow and light-green region in Fig.~\ref{fig2}. 
Taking into account the rule found in~\citep{Cierniak:2020eyh}, that the maximum mass of the hybrid star branch can be situated up to $0.2~M_\odot$ above the SP, one finds the upper limit for the maximum mass of hybrid stars (indicated by the green dashed line in Fig.~\ref{fig2}) and the lower limit of the SP mass 
(indicated by the light-green region in Fig.~\ref{fig2}) that would be compatible with the maximum mass constraint from the mass measurement on PSR J0740+6620~\citep{Cromartie:2019kug}.

%%%%%%%%%%%%%%%%%%%%%%%%%%%%%%%%%%%%%%%%%%%%%%%%%%%%%%%%

As an alternative to the DD2 class of EoS, the second class of  hadronic EoS considered here is based on the Argonne V18  nucleon--nucleon potential and three-body force together with a repulsive multi-pomeron interaction~\citep{Yamamoto:2017wre}. 
The strength of the latter lies in its ability to accurately describe the differential scattering cross--section measured for $^{16}$O-$^{16}$O collisons at $E_{\rm Lab}=70$ MeV also at large angle, as well as reproducing the nuclear saturation properties together with a high density stiffening of the EoS that is needed to solve the famous "hyperon puzzle" problem. 

In Fig.~\ref{fig2} we show by solid lines the $M-R$ relations for three examples of this class of EoS which represents the softest "reasonable" hadronic EoS model which may therefore serve as an estimator for the smallest neutron star radii that are described with a purely hadronic EoS. 
The dash-dotted lines (including also the APR EoS) correspond to EoS without hyperons and are therefore not realistic above the hyperon onset mass of about $1.5~M_\odot$ where they may be ignored.
This limiting radius is well approximated by the segment of the black dotted line labelled $c_s^2=0.7$ in Fig.~\ref{fig2} that lies in the region of the mass for PSR J0740+6620.
Note that among typical RDF EoS with hyperons fulfilling the maximum mass constraint $M_{\rm max}\ge 2.0~M_\odot$ the lowest radius at the maximum mass is 10.5 km, see~\citep{Fortin:2014mya,Maslov:2015wba,Raduta:2019rsk}.

\begin{figure}[!thb]
 \centering
\includegraphics[width=0.45\textwidth]{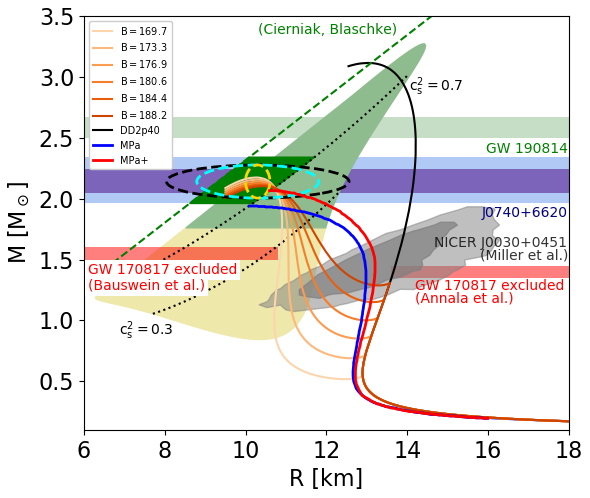} 
\caption{Same as Fig.~\ref{fig2}, but with hybrid star sequences and fictitious results of a NICER radius measurement on PSR J0740+6620 that could only be explained by a hybrid NS, but not by a purely hadronic NS interior. The hybrid NS sequences correspond to $c_s^2=0.7$ and $A=175.9$ MeV/fm$^3$,  for different masses at the onset of deconfinement, which correspond to the values of the bag pressure $B$ (in MeV/fm$^3$) given in the legend, see the black curve in Fig.~\ref{fig4}.  For details, see text.}
\label{fig3}
\end{figure}

Consequently, the rhombic region in the $2~M_\odot$ mass range that is highlighted in dark-green color in Fig.~\ref{fig2} can be accessed only by hybrid stars. 
Examples for hybrid star sequences that would populate this region are shown in Fig.~\ref{fig3}.
We note that all of them lie on a third family 
branch~\citep{Gerlach:1968zz,Blaschke:2013ana,Alvarez-Castillo:2018pve,Blaschke:2020qqj}
and thus imply the phenomenon of mass twin stars~\citep{Glendenning:1998ag,Benic:2014jia,Alvarez-Castillo:2017qki} 
at its low-mass end.
A measurement of mass and radius for a pulsar that falls in this region would allow the conclusion that this pulsar harbors a core made of deconfined quark matter~\citep{Cierniak:2020eyh}. 

The presently ongoing NICER measurement of the radius for PSR J0740+6620 has the potential to result in such a discovery!
For this pulsar there is a very precise, uncorrelated mass determination based on the Shapiro-delay measurement
which resulted in $M_{J0740}=2.14^{+0.10}_{-0.09}~M_\odot$  \citep{Cromartie:2019kug}, that is even subject to improvement. 
The NICER radius measurement for that pulsar is not as good as for PSR J0030+0451 due to a lower X-ray count rate.
Before the results of that actual measurement will be released, we make assumptions for the radius uncertainty being 
1.5, 1.0 and 0.3 km.  
Then, by setting the median of $R_{J0740}$ to 10.3 km in the center of the dark-green rhombic region, we see in Fig.~\ref{fig3} that only for the smallest hypothetical standard deviation the confined quark hypothesis could be excluded.

%with the minimal radius for hyperonic NS at $2.14~M_\odot$ being 11 km, we obtain that for those three assumptions for 
%the $1\sigma$ standard deviation in the radius measurement the confined quark hypothesis could
%be excluded if the median of $R_{J0740}$ were less than 9.5, 10.0 and 10.5 km, respectively.

%If we estimate the $1\sigma$ error region of~\citep{Miller:2019cac} by a tilted ellipse~\citep{Blaschke:2020qqj} then this conclusion would be possible when at a mass of $2.1~M_\odot$ the radius of this pulsar would be $\le 9.7$ km. 
%Would the standard deviation be lower by a factor $2.5$, the radius should be  $\le 10.5$ km, see the dashed-line ellipses in Fig~\ref{fig3}. 

A measurement with this median and even the largest assumed standard deviation, however, would exclude the "two-families" scenario~\citep{Drago:2020gqn} according to which stars with large masses $M\ge 2~M_\odot$ should be strange stars with radii $\ge 14$ km while the maximum mass of the smaller hadronic stars stays well below the $2~M_\odot$ limit.  
%%%%%%%%%%%%%%%%%%%%%%%%%%%%%%%%%%%%%%%

\subsection{Onset of deconfinement vs. $M_{\rm max}$}

\begin{figure}[htb]
\centering
\includegraphics[width=0.45\textwidth]{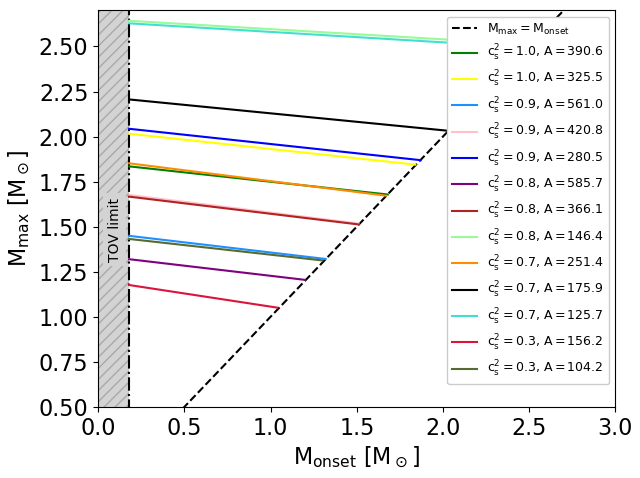} 
\caption{Maximum mass of hybrid stars vs. onset mass of the Maxwell-constructed deconfinement transition from hadronic matter described by DD2p60 EoS and  CSS quark matter. For fixed position of the SP, determined by the values of the squared speed of sound $c_s^2$ and the slope parameter $A$ (in MeV/fm$^3$), the onset mass depends on the bag pressure $B$. Lowering the onset mass increases the maximum mass according to Eq.(\ref{eq:lever}). 
At fixed $c_s^2$ a lower slope parameter increases the stiffness and thus the maximum mass of the EoS.
} 
\label{fig4}
\end{figure}

We have mentioned in the previous section that the maximum mass on a hybrid star branch lies up to $\sim 0.2~M_\odot$ above that of the SP. 
In the opposite extreme, the mass of the SP $M_{\rm SP}$ may be equal to the maximum mass, namely when the onset mass for deconfinement is very high itself. 
Consequently, there is a relation like a "lever rule", that for a given SP position characterised by the values of $c_s^2$ and $A$ in the quark matter EoS (\ref{css3}) the maximum mass increases when the onset mass decreases, see Fig.~\ref{fig4}. 
The results of Fig.~\ref{fig4} can be described by a linear fit formula
\begin{equation}
\label{eq:lever}
M_{\rm max}=M_{\rm SP} + \delta (M_{\rm SP} - M_{\rm onset})~,
\end{equation}
where $\delta=0.1$ and each curve in that figure corresponds to a given value of $M_{\rm SP}=M_{\rm SP}(c_s^2,A)={\rm const}$.
This explains the observed degeneracy of the lines: a change in $c_s^2$ is compensated by a choice of the slope parameter $A$ so that $M_{\rm SP}$ remains unchanged.

We note that for $M_{\rm onset}\gtrsim 1.4~M_\odot$ the lines in Fig.~\ref{fig4} should be regarded with caution because hyperons are missing in the DD2 class of hadronic EoS and for models with larger excluded nucleon volume (DD2p40, DD2p60, DD2p80) the radius constraint of \citep{Annala:2017llu} deduced from the tidal deformabilities in the mass range of GW170817 is violated.

\begin{figure}[htb]
\centering
\includegraphics[width=0.45\textwidth]{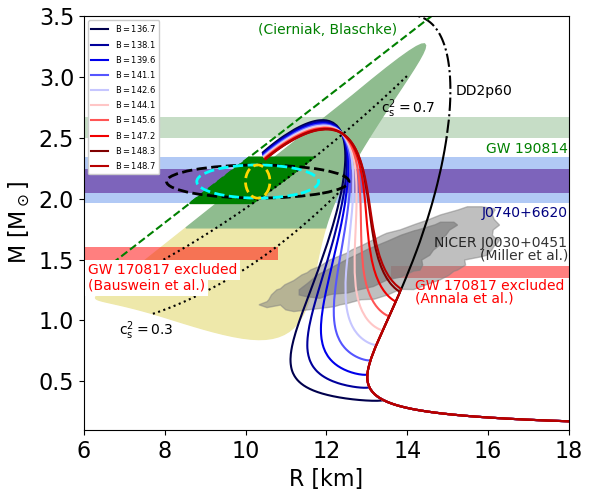} 
\caption{Mass-radius diagram for a hybrid star family that corresponding to the DD2p60-CSS hybrid EoS introduced in the text.
The maximum mass on the hybrid star branch reaches the mass range of the lighter compact object in GW190814~\citep{Abbott:2020khf}, 
which therefore could be a hypermassive neutron star. 
The sequences with lowest onset masses for deconfinement exhibit the twin phenomenon.} 
\label{fig5}
\end{figure}

In table \ref{table} we list the densities in the center of the neutron stars with the mass $M_{\rm onset}$ at the onset of quark deconfinement. 
\begin{table}
\caption{Central densities $n_c$ of neutron stars with the mass $M_{\rm onset}$ at the onset of quark deconfinement for the RDF EoS DD2 with different excluded volumina, $n_0=0.15$ fm$^{-3}$.
}
\label{table}
\begin{tabular}{c|c|c|c|c}
 & \multicolumn{4}{c}{$n_c[n_0]$}\\
$M_{\rm onset}[M_\odot]$ & DD2p0 & DD2p40 & DD2p60 & DD2p80 \\
\hline
0.263 & 1.107 & 1.088 & 1.073 & 1.063\\
0.435 & 1.365 & 1.264 & 1.206 & 1.162\\
0.626 & 1.600 & 1.426 & 1.328 & 1.255 \\
0.833 & 1.814 & 1.570 & 1.437 & 1.339 \\
1.028 & 1.993 & 1.686 & 1.523 & 1.406\\
1.218 & 2.166 & 1.787 & 1.597 & 1.464 \\
1.385 & 2.327 & 1.874 & 1.659 & 1.511 \\
1.553 & 2.512 & 1.963 & 1.722 & 1.559\\
1.699 & 2.698 & 2.047 & 1.779 & 1.602\\
1.832 & 2.897 & 2.132 & 1.838 & 1.644  
\end{tabular}
\end{table}
In our applications shown in Figs. \ref{fig3} and \ref{fig5}, we consider hybrid EoS with the RDF models DD2p40 and DD2p60 for the hadronic phase without hyperons and with an early onset of deconfinement at $M_{\rm onset}\lesssim 1.5~M_\odot$.
It is illustrated in Fig.~\ref{fig2} that for EoS even softer than DD2p40 the hyperon onset is at a mass $\sim 1.5~M_\odot$ and corresponds to a threshold density for hypernuclear matter at $\sim 2~n_0$ \citep{Fortin:2014mya,Maslov:2015wba,Raduta:2019rsk}.  
The neglect of hyperons in the present work is thus acceptable since for the stiffer RDF models the central densities for stars
with $M\lesssim 1.5~M_\odot$ are below the hyperon threshold, see  table \ref{table}.
Thus, our approach falls into the class of models where the absence of hyperons is justified by the early onset of deconfinement before the hyperon threshold could be reached in the hadronic phase of the matter. 
Such a scenario is the favorable solution of the so-called hyperon puzzle \citep{Baldo:2003vx,Zdunik:2012dj} which has been 
revisited under modern mass and radius constraints in \citep{Shahrbaf:2019vtf}.
For a recent review on the NS EoS, see  \citep{Blaschke:2018mqw} and references therein.

From Fig.~\ref{fig4} we can read off another remarkable feature of the class of hybrid EoS discussed here. 
It can describe hypermassive neutron stars with $M>2.6~M_\odot$ as hybrid stars with a quark matter core!
Moreover, these hypermassive hybrid stars belong to a third family branch that exhibits the twin star property as shown in Fig.~\ref{fig5}.
Thus we have demonstrated that the upper limit for the maximum mass of hybrid star sequences that still allow
twins could be much higher than recently conjectured by~\citep{Christian:2020xwz}, where it was claimed that the existence of neutron stars with masses 
$M>2.2~M_\odot$ would rule out twin stars. 
With our result shown in Fig.~\ref{fig5} we disprove this claim.

Another interesting result follows from Fig.~\ref{fig5}, namely that the lighter compact object in the binary merger GW190814 with the mass of 
$2.5$ - $2.67~M_\odot$~\citep{Abbott:2020khf} can be a hybrid neutron star with a quark matter core.  
It may even belong to a third-family branch of hybrid stars.
For a discussion of EoS constraints when the companion star of GW190814 being a neutron star and not a black hole, see also 
\citep{Kanakis-Pegios:2020kzp,Tsokaros:2020hli}.
It is not possible from GW190814 alone to conclude whether the lighter companion was the lightest black hole or the heaviest neutron star observed to date.
But following~\citep{Vattis:2020iuz} it is not likely that this object could have been a primordial black hole. 
Its mass is, however, well in the range that could be expected from a merger of two typical-mass neutron stars. 
See, e.g.,~\citep{Bauswein:2020aag} for extracting the maximum NS mass from multi-messenger observations of binary NS mergers.
If the neutron star interpretation shall become favorable, more compact stars with a mass of $\sim 2.6~M_\odot$ shall become observable.
Then it is likely that the merger GW170817 could have resulted in a long-lived pulsar that powers the afterglow of the kilonova event, see~\citep{Troja:2020pzf}.
%%%%%%%%%%%%%%%%%%

Finally, the question of a possible very early onset of deconfinement with $M_{\rm onset}\lesssim 1~M_\odot$ could be settled with a precise radius measurement 
for the lightest neutron stars. A candidate for a NICER radius measurement could be the millisecond pulsar J0751+1807 with a mass of $1.26\pm0.14~M_\odot$
and a period $P=3.48$ ms \citep{2007AAS...21110206N}. 
Comparing Fig.~\ref{fig5} and  Fig.~\ref{fig2} for the radii in the 1$\sigma$ mass range covered by J0751+1807, a measurement of a radius less than $\sim 13$ km 
(as suggested by the recent analysis of GW170817 in~\citep{Capano:2019eae} which resulted in $R_{1.4M_\odot}=11.0^{+0.9}_{-0.6}$ km) 
would exclude the hadronic NS interpretation since the softest realistic hadronic EoS corresponds to radii of $\sim 13$ km, see Fig.~\ref{fig2}.
As can be seen in Fig.~\ref{fig5}, a radius $R_{1.4M_\odot}<12.5$ km is well explained by the high-mass hybrid star scenario with an onset of deconfinement at $M_{\rm onset} <1.0~M_\odot$.
Also in the case when the NICER radius measurement for PSR J0740+6620 would yield $R_{2.14~M_\odot}\lesssim11$ km and thus render the hypermassive (hybrid) neutron star scenario for GW190814 unlikely, the radius constraint of ~\citep{Capano:2019eae} entails an onset of deconfinement at a subsolar mass, see Fig.~\ref{fig1}. 
A NICER radius measurement for a low-mass pulsar like PSR J0751+1807 could bring an independent confirmation for such a claim.
In this case the two NS that merged in the event GW170817 would both have been hybrid stars with an extended quark matter core, see also~\citep{Blaschke:2020qqj}.
Recently, it has been confirmed that the signal for a first-order phase transition in the frequency spectrum of the postmerger gravitational waves suggested in~\citep{Bauswein:2018bma} would also be applicable when the merging NS are hybrid stars ~\citep{Bauswein:2020ggy}.

\section{Conclusions}

In this work we have used the property of hybrid NS sequences to possess a special point in the $M-R$ diagram in order to investigate the consequences that can be drawn for the onset of deconfinement. To this end we have employed a generic three-parameter CSS EoS for the description of the quark matter phase.
A key observation is that the APR EoS has to be excluded as a candidate for the softest realistic hadronic EoS that could be used to estimate the lowest possible hadronic NS radii because it has a severe hyperon puzzle problem.
Instead the EoS of~\citep{Yamamoto:2017wre} is suggested here to be the suitable one, yielding minimal hadronic NS radii which allow to conclude for the existence of a core of deconfined quark matter when $R_{2.14~M_\odot}<11$ km and/or $R_{1.4M_\odot}<12.5$ km. 
The latter result, already supported by the refined analysis of~\citep{Capano:2019eae} of the multi-messenger observations of the merger GW170817, would not only support an early onset of quark deconfinement but also the scenario that the $2.6~M_\odot$ object in the binary merger GW190814 could have been a hypermassive hybrid neutron star and not a black hole.
This conclusion, however, could be falsified when the NICER radius measurement of the massive pulsar PSR J0740+6620 would result in a value $R_{2.14~M_\odot}<11$ km. Then, this object would be identified as a hybrid NS with a large quark matter core, but at the same time the hypermassive (hybrid) NS scenario for the lighter object in the binary merger GW190814 would be excluded and one could conclude that this event was a binary black hole merger.
One can look forward to the next binary NS merger event which could then confirm the scenario of an early onset of quark deconfinement in NS provided that not only the tidal deformability of the inspiral phase but also the peak frequency of the postmerger gravitational wave signal could be measured~\citep{Bauswein:2018bma}.

\subsection*{Acknowledgements}
We would like to thank Andreas Bauswein, Cole Miller, Tom Rijken and Yasuo Yamamoto for their comments on this paper.
D.B. acknowledges support from the Polish National Science Center under grant No. 2019/33/B/ST9/03059 
and from the Russian Foundation for Basic Research under grant No. 18-02-40137.

%\bibliography{David}

\end{document}